# A Structurally Self-Assembled Peptide Nano-Architecture by One-Step Electrospinning

Robabeh Gharaei,[†,§] Giuseppe Tronci,[†,‡] Robert P.W. Davies,[‡] Caroline Gough,[†] Reem Alazragi,[⊥] Parikshit Goswami,[§] and Stephen J. Russell [†]

[†]Nonwovens Research Group, School of Design, University of Leeds, Leeds LS2 9JT, United Kingdom

[§]Fibre and Fabric Functionalisation Research Group, School of Design, University of Leeds, Leeds LS2 9JT, United Kingdom

[‡]Biomaterials and Tissue Engineering Research Group, School of Dentistry, St. James's University Hospital, University of Leeds, Leeds LS9 7TF, United Kingdom

[†]Division of Oral Biology, School of Dentistry, University of Leeds, Leeds LS2 9JT, United Kingdom

[⊥]Centre for Self-Organising Molecular Systems, School of Chemistry, University of Leeds, Leeds LS2 9JT, United Kingdom

**Table of contents entry**

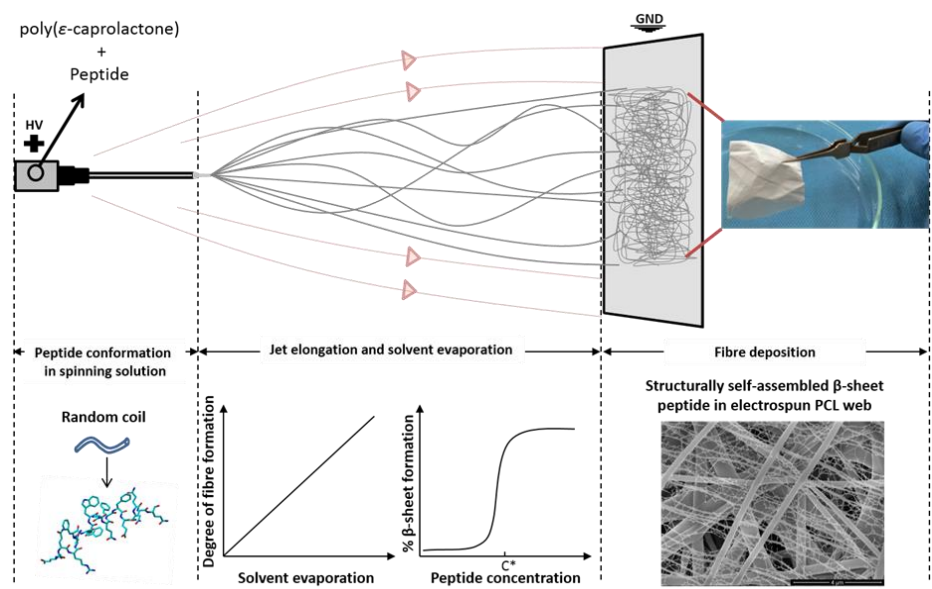

Peptide self-assembly during electrospinning while the solvent is evaporating and the fibres are forming.

**Abstract**

Self-assembling peptides (SAPs) have shown to offer great promise in therapeutics and have the ability to undergo self-assembly and form ordered nanostructures. However SAP gels are often associated with inherent weak and transient mechanical properties and incorporation of them into polymeric matrices is a route to enhance their mechanical stability. The aim of this work was to incorporate $P_{11}$-8 peptide (CH3COQQRFOWOFEQQNH2) within poly($\varepsilon$-caprolactone) (PCL) fibrous webs via one-step electrospinning, aiming to establish the underlying relationships between spinning process, molecular peptide conformation, and material internal architecture. Electrospinning of PCL solutions (6% w/w) in hexafluoro-2-propanol (HFIP) containing up to 40 mg·mL$^{-1}$ $P_{11}$-8 resulted in the formation of fibres in both nano- (10-100 nm) and submicron range (100-700 nm), in contrast to PCL only webs, which displayed a predominantly submicron fibre distribution. FTIR and CD spectroscopy on both PCL/peptide solutions and resulting electrospun webs revealed monomeric and β-sheet secondary conformation, respectively, suggesting the occurrence of peptide self-assembly during electrospinning due to solvent evaporation. The peptide concentration (0 → 40 mg·mL$^{-1}$) was found to primarily affect the internal structure of the fabric at the nano-scale, whilst water as well as cell culture medium contact angles were dramatically decreased. Nearly no cytotoxic response (> 90% cell viability) was observed when L929 mouse fibroblasts were cultured in contact with electrospun peptide loaded samples. This novel nanofibrous architecture may be the basis for an interesting material platform for e.g. hard tissue repair, in light of the presence of the self-assembled $P_{11}$-8 in the PCL fibrous structure.

**Introduction**

Peptide self-assembly has been the focus of research in the last two decades in light of their potential application in hard and soft tissue repair as well as in controlled drug delivery.[1-4]

Rationally designed SAPs are composed of amino acidic building blocks that can mimic specific molecular features of the extracellular matrix (ECM) of tissues, such as the RGD cell-binding peptide sequence,[2] and facilitate the enhancement of cell growth in the biomaterial.[3, 5, 6] Designing different primary peptide structures by applying various amino acid side chains and altering peptide sequences, enables the physical and biological properties of peptides to be tuned according to the intended end use.[2]

The 11-residue family of peptides ($P_{11}$-X) consists of negatively or positively charged hydrogelating materials and self-assemble hierarchically into long β-sheet tapes (a single-molecule thick), ribbons (two stacked tapes), fibrils (multiple stacks of ribbons) and entwined fibrils (referred to by some as fibres), following application of external stimuli.[6] Above a critical concentration (c*), peptide monomers assemble into hydrogen bonded β-sheet tapes, with higher order structures being produced if the concentration is further increased. This class of peptides also undergoes pH[7] and ionic strength-triggered[8] self-assembly which is relevant for their applicability as drug delivery vehicles. So far, most of the published studies using $P_{11}$-X peptides have focused on the molecular design of self-supporting gels, whereby promising properties have been shown with regard to cell growth and hard tissue deposition.[2, 3, 6, 9] Furthermore peptide $P_{11}$-8 (+2 charge) has shown low immunogenicity *in vivo*,[10] no cytotoxic effect to human and murine cells[6, 11] and enhanced bone tissue regeneration. However, despite their inherent biofunctionality, self-assembled peptide gels frequently suffer from poor mechanical strength and lack of structural stability. This potentially makes their handling and fixation during implantation *in vivo* challenging, particularly in large load-bearing tissue defects. Attaining mechanically-competent scaffolds capable of supporting cell growth spatially and temporally until the newly engineered tissue is formed, is one of the major challenges in the design of SAP-based medical devices.[12] The incorporation of SAPs within water-stable synthetic fibres is a potential route to address this challenge, and to deliver structurally reinforced fabrics.

Electrospinning is an established technique for the production of polymeric nanofibres, the fundamentals of which have been known since the work of Formhals in the 1930s.[13] Electrospun webs have high solid surface area to volume ratio and high porosity and control of fibre morphological and topographical features is readily achieved by varying process parameters.[14] Their application as tissue scaffolds has been extensively studied.[13, 15-18] In addition to the structural features that can mimic fibrous architecture of the ECM[19], the material composition should be nontoxic and induce the intended response on tissue components *in vivo* or *in vitro*.[16]

A variety of synthetic and natural polymers alone or in blends has been successfully electrospun resulting in fibres with diameters ranging from the micron up to the nanometre scale,[5, 16, 20-26] whilst voltage-free spinning techniques have also been recently employed for the formation of single biomimetic fibres.[27, 28] Biocompatible synthetic polymers that have been successfully electrospun include aliphatic linear polyesters such as polyglycolic acid (PGA),[29] polylactic acid (PLA)[30, 31] and PCL.[32, 33] Among these, PCL can be used in FDA approved devices i.e. for applications in the human body such as drug delivery and tissue engineering.[32, 34] PCL is low-cost, non-toxic and biodegrades slowly depending on the molecular weight. PCL has been extensively investigated in relation to various biomedical applications, including regenerative therapies,[23, 25, 35-37] whilst showing supporting the attachment and growth of muscle cells, mesenchymal stem cells, and chondrocytes.[32, 38-40] However, due to the lack of cell-binding sequences along its polymeric backbone and its hydrophobicity, PCL must be modified via e.g. plasma treatment and/or coating.[32,41,42]

In addition to synthetic polymers, the self-assembly of some (poly-)peptides and proteins was exploited using electrospinning, aiming to accomplish biomimetic and unusual architectures. Nuansing et al.[43] attempted to electrospin a short peptide of Fmoc-FG (Fmoc–Phe–Gly) which resulted in rough fibres with diameters of around 300 nm, while Haynie et al.[15] successfully electrospun polypeptides of poly(L-ornithine) (PLO) and poly(L-glutamic acid4-co-L-tyrosine) (PLEY) resulting in fibres with diameters ranging from 0.1µm to several microns. Tayi et al.[44] also attempted to electrospin peptide amphiphiles (PAs) into micrometre-scale fibres without a polymer carrier. In all of these studies, in addition to the serious challenges involving the optimisation of the spinning solution as well as electrospinning process parameters, the resulting pure electrospun (poly)- peptides webs were not mechanically robust such that they are not easy to handle for clinical use. Since the self-assembly is only stabilised by weak non-covalent bonds, resulting architectures can only partially be customised, so that electrospinning of classical synthetic polymers, e.g. PCL, has been studied in the presence of bioactive SAPs. Andukuri et al.[41] coated electrospun PCL fibres with peptide amphiphiles (PAs) for cardiovascular implants, whereby PAs were self-assembled at the surface of the nanofibres by a solvent evaporation technique. Significant improvement was observed in adhesion and proliferation of human umbilical vein endothelial cells (HUVECs), however there were limitations in smooth muscle cell proliferation and adhesion of platelet cells.[41] In another study, electrospun PCL scaffold was covalently modified by perlecan domain IV peptide, and this approach led to significantly enhanced cell adherence and infiltration in a 3-D pharmacokinetic cancer model.[42] To avoid the post-treatments, the biofunctional components can also be directly blended with the polymer in solutions to fabricate electrospun fibres in one step. A single step electrospinning of mixture of poly(lactic-co-glycolic acid) (PLGA) and peptide (CGGRGDS) has been demonstrated by Gentsch et al.[45], which resulted in fibre surface enrichment with peptide. In other approaches, PCL/peptide conjugates were electrospun to specifically and non-covalently guide the spatial arrangement of biomolecules such as glycosaminoglycans (GAGs) within the scaffold, so that biological gradients found in native tissues could be successfully mimicked.[46] Further to that, a variety of self-assembling peptides (EAK, DAK, EAbuK, EYK, RGD-EAK, and RGD-EAKsc) have been added into PCL solutions prior to electrospinning.[20] The resulting fibres containing peptide not only exhibited higher surface wettability and amorphous phase compared to that of PCL, but there was also an improvement in *h-osteoblast* cell adhesion.

The aim of this work was to investigate whether a structurally stable fibrous architecture could be accomplished in one-step via (i) electrospinning of a HFIP solution of both peptide $P_{11}$-8 and PCL and (ii) molecular peptide self-assembly during fibre formation. $P_{11}$-8 peptide was added into a HFIP solution of PCL. HFIP was selected as electrospinning solvent in light of its fast evaporation and ability to break down hydrophobic interactions and hydrogen bonds,[47] so that clear solutions of PCL and $P_{11}$-8 random coils could be obtained. Electrospun webs were inspected by SEM and TEM, revealing the formation of both submicron (diameter of 100-700 nm) and nanofibres (diameter of 10-100 nm), whilst homogeneously distributed submicron fibres were observed in the case of PCL web controls. The secondary conformation of peptide before and after electrospinning and the mechanism behind the formation of a secondary peptide concentration-dependent nano-architecture in PCL/$P_{11}$-8 electrospun webs was elucidated via spectroscopic analysis. Resulting fibres were then cultured with L929 mouse fibroblasts and cell response were analysed in terms of cytotoxicity and metabolic activity.

**Experimental Methods**

**Preparation of PCL and $P_{11}$-8 spinning solution.** PCL ($M_n$: 80,000 g.mol$^{-1}$) and HFIP (purity ≥ 99.0%) were purchased from Sigma Aldrich UK. Peptide $P_{11}$-8 (CH$_3$CO-Gln-Gln-Arg-Phe-Orn-Trp-Orn-Phe-Glu-Gln-Gln-NH$_2$ (peptide content ~ 75%, HPLC purity of 96%) was purchased from CS Bio Co. (USA). The peptide product dry weight reflects 25% non-peptide content, ascribed to residual amount of water and trifluoroacetic acid (TFA) counterions bound to positively charged groups. Fluorescein-β-Ala–$P_{11}$-4 (Fluorescein-β-Ala-Gln-Gln-Arg-Phe-Glu-Trp-Glu-Phe-Glu-Gln-Gln-NH$_2$ (HPLC purity 95.1%) was purchased from PolyPeptide (France). The electrospinning solutions were prepared by dissolving 6% (w/w) PCL in HFIP at room temperature. After magnetic stirring for 24 h, either 10, 20 or 40 mg of $P_{11}$-8 was added into 1 mL of the PCL solution and they were left for 24 h until homogeneous mixtures were obtained. A control solution of only PCL in HFIP (6% w/w) was also prepared.

**Preparation of PCL and $P_{11}$-8 fibre webs via electrospinning.** A standard single spinneret electrospinning setup consisting of a syringe connected to high voltage power supply (Glassman Inc.), a grounded collector and a syringe pump (Kd Scientific Model 200 Series) was used in this study. The solutions were released from a 5 mL glass syringe fitted with a 22 gauge blunt tipped needle (Sigma Aldrich) at a rate of 1 mL.h$^{-1}$. Electrospinning was carried out at a distance between needle and collector of 18 cm with a voltage of 20 kV, enabling the formation of smooth and uniform fibres with no beads. An ambient temperature of 21 ± 2°C and a relative humidity of 43 ± 2% were consistently recorded in the spinning chamber. Fibres were collected on an aluminium foil of 15×15 cm and after deposition they were dried in room temperature under vacuum for 7 days to ensure evaporation of all solvent residues. Sample nomenclature is defined in Table 1.

**Table 1.** Sample nomenclature and formulation used in this study. All samples were prepared from a HFIP solution of PCL (6% w/w).

| Sample ID | Sample specification |
|---|---|
| 1 | PCL ([$P_{11}$-8]: 0 mg.mL$^{-1}$) |
| 2 | PCL/$P_{11}$-8 ([$P_{11}$-8]: 10 mg.mL$^{-1}$) |
| 3 | PCL/$P_{11}$-8 ([$P_{11}$-8]: 20 mg.mL$^{-1}$) |
| 4 | PCL/$P_{11}$-8 ([$P_{11}$-8]: 40 mg.mL$^{-1}$) |
| 5 | PCL/$P_{11}$-8/fluoro $P_{11}$-4 ([$P_{11}$-8]: 20 mg.mL$^{-1}$ and [fluoro $P_{11}$-4]: 330 µg.mL$^{-1}$) |

**CD & FT-IR Spectroscopy.** FTIR and CD were used to determine the secondary structure of the peptide in both solution and fibre state.[48, 49] The solution for CD and FTIR analysis was prepared in a similar manner to that used for electrospinning by dissolving 20 mg of $P_{11}$-8 in 1 mL solution of PCL in HFIP (6% w/w) (sample 3). FTIR of the solution was performed with a Nicolet 6700 FTIR. The sample was held onto CaF$_2$ windows and 32 scans were performed and measurements taken in the range 4000 – 400 cm$^{-1}$. The HFIP solvent spectrum (blank) was subtracted from the spinning solution spectrum. CD spectra were recorded using a Chirascan CD spectrometer (Applied

Photophysics) and the solutions were analysed in 1 mm path length cuvettes at 22°C (spinning conditions). The data was acquired at a step resolution of 1 nm with a scan speed of 60 nm.min$^{-1}$ and a bandwidth of 4.3 nm. Far UV spectra were recorded over the wavelength range 190 to 240 nm. Each spectrum shown herein was the average of two scans. The HFIP (blank) spectrum was subtracted from the peptide-containing spinning solution spectrum. The data then were converted to mean residue molar ellipticity (deg.cm$^2$.dmol$^{-1}$) and finally a 7$^{th}$ order polynomial fitting was performed ($R^2$= 0.95). Infrared spectra of PCL/P$_{11}$-8 fibres (sample 3) deposited on aluminium foil were obtained using a Perkin Elmer FTIR with diamond Attenuated Total Reflectance (ATR) attachment system. A total of 64 scans were averaged for each spectrum in the range between 4000 and 400 cm$^{-1}$.

**Electron Microscopy (SEM & TEM).** Dry electrospun samples were sputter coated with platinum with a thickness of 8 nm and imaged using a field emission gun scanning electron microscope (LEO1530 Gemini). The microscope was provided with an energy-dispersive X-ray spectrometer (EDX) of Oxford Instruments AztecEnergy to investigate the chemical composition of the as-spun materials. To more extensively investigate the nano- and submicron architecture, a thin layer of the fibres were directly electrospun onto transmission electron microscopy (TEM) grids (mounted on aluminium foil) and TEM analysis was performed using a FEI Tecnai TF20. The TEM microscope was also fitted with an EDX system (Oxford Instruments INCA 350). Three random spots per sample was tested for EDX spectroscopy. PCL and PCL/P$_{11}$-8 electrospun fibre dimensions (sample 1 and 3) were measured using multiple SEM images by Image Pro Plus 6.2.1 software (Media Cybernetics) with at least 100 measurements per sample to determine mean fibre diameter and associated frequency distributions. Values were expressed as mean ± standard error.

**Confocal laser scanning microscopy (CLSM).** To investigate the peptide distribution throughout the fibres, the P$_{11}$-8 peptide in the spinning solution was doped with P$_{11}$-4 functionalised by a fluorescein moiety [ratio of fluoro P$_{11}$-4:P$_{11}$-8 = 1:60] to facilitate viewing by confocal microscopy. Functionalised $P_{11}$-4 was used in this case to produce a polyelectrolyte β-sheet complex. Mixing an anionic and cationic peptide is an alternative way to induce self-assembly,[50] which in this case was the preferred method to ensure full mixing of the fluoro tagged peptide. A large fluorescein moiety will have a high degree of steric hindrance which would, in an anionic – anionic system phase separate and exclude itself from the bulk of the β-sheet structures formed. Both the blank (sample 1) and fluoro $P_{11}$-4 doped peptide-loaded sample (sample 5) were observed by confocal laser scanning microscopy of Zeiss LSM510.

**Surface Wettability.** PCL and PCL/$P_{11}$-8 fibres and cast films were tested for contact angle measurement using goniometry (FTA 4000 Microdrop®). 1.5 $\mu$L drop of either deionised water (Milli-Q) or Dulbecco's Modified Eagle's medium (DMEM) was deposited and the contact angles over 15 s were measured. Three replicates per sample were examined. Values were expressed as mean ± standard error. Films were also prepared to investigate the surface wettability of the non-porous material. PCL and PCL/$P_{11}$-8 solutions in HFIP were cast on glass slides and films were prepared by drying under vacuum for 48 hours. Contact angle measurements were carried out with deionised water and results expressed as described above.

**Direct Cytotoxicity assay.** Fibres were prepared in the same manner as previously described and sterilised by gamma irradiation. Samples of 10×10 mm were cut out for evaluation. PCL (sample 1) was used as negative control because of its non-cytotoxicity[32, 37] and the PCL/ $P_{11}$-8 (sample 3) was used to test the cytotoxic response of $P_{11}$-8 directly in contact with L929 cells. L929 mouse fibroblast cells were cultured in DMEM (Dulbecco's modified eagle's medium) with 10% foetal bovine serum and 1% Penicillin/Streptomycin (P/S) until confluence (5% $CO_2$, 37°C). Gamma-sterilised, DMEM-soaked samples were singularly placed on the wells of a 24-well plate in which cells with a seeding density of $10^5$ cells.mL$^{-1}$ had been cultured for 24 h with verified sub-confluency (n=6). The culture medium was then replaced in each well with 500 μL of fresh

DMEM. Alternatively 250 μL DMEM and 250 μL DMSO (Dimethyl sulfoxide, purity≥ 99.9) was added in positive control wells. Plates were incubated for 48 h at 37ºC in 5% (v/v) $CO_2$ in air. The PCL and PCL/$P_{11}$-8 samples in contact with cells were examined under a phase-contrast microscope (Leica DFC365 FX). After aspirating the culture medium and washing the samples with PBS, 200 μL of filter sterilised MTT solution (1 mg.mL$^{-1}$ of (3-(4, 5-dimethylthiazol-2-yl)-2,5-diphenyltetrazoliumbromid) was added to each well and the plates were incubated at 37 ºC for 2 h. Then the MTT solution was removed from each well and replaced with 400 μL isopropanol to dissolve the generated blue-violet insoluble formazan. After swaying, the plates were placed in a micro-plate reader and readings were obtained at 570 nm and 650 nm. The quantity of formazan product can be measured by the following equation (Equation 1), which is directly proportional to the number of living cells in each culture:

$OD = OD_{570} - OD_{65}$ **(Equation 1)**

where $OD_{570}$ and $OD_{650}$ are the mean values of the measured optical density of the test sample at 570 and 650 nm, respectively.

**Results and Discussion**

Electrospun webs of PCL and $P_{11}$-8 were successfully produced and the nano- and microscale structures were studied to elucidate morphology, chemical composition, peptide distribution, molecular peptide conformation, wettability and cytotoxic response.

**Secondary conformation of peptide.** It was intended to incorporate the peptide within the spinning solution in its monomeric form, such that changes in concentration during electrospinning and solvent evaporation could potentially elicit β-sheet formation. CD and FTIR analysis were conducted to identify the conformation of $P_{11}$-8 before (in solution) and after electrospinning (in as-spun fibres) (Figure 1). The CD spectrum of the $P_{11}$-8 solution prior to spinning displays a negative minimum at around 195 nm and a positive maximum at around 210 nm (Figure 1A), which is consistent with the random coil peptide conformation in the solution reported previously.[51] This explanation is supported by the lack of a negative band at 218 nm, which would be characteristic of a β-sheet conformation.[6, 48] Additionally, a slight shift in wavelength and intensity of the peptide bands was observed when comparing the obtained spectra with previously reported CD plots.[6, 48]

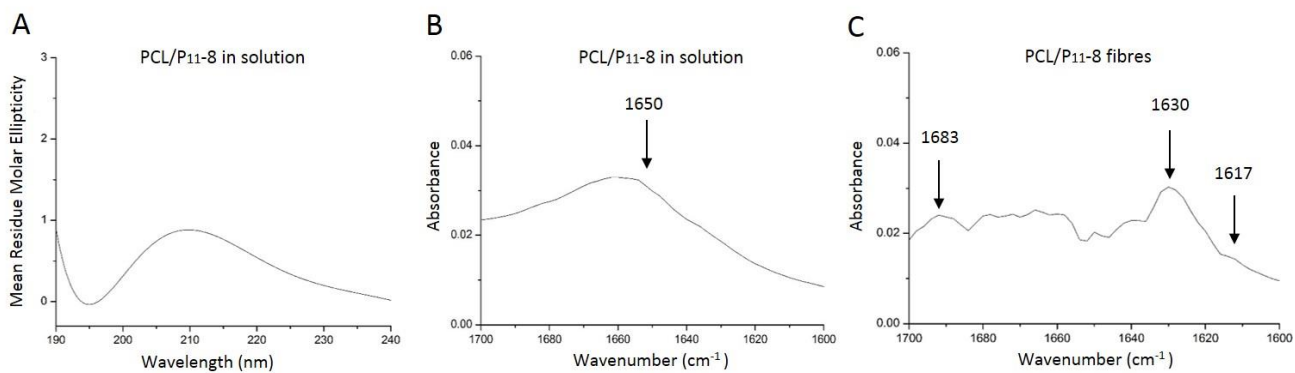

**Figure 1.** Secondary structure analysis of $P_{11}$-8 before and after electrospinning in the solution and in the as-spun fibres. (A) CD spectrum of PCL/$P_{11}$-8 solution in HFIP, (B) FTIR amide I′ bands of electrospinning solution of PCL/$P_{11}$-8 in HFIP, (C) ATR-FTIR amide I′ bands of electrospun fibres of PCL/$P_{11}$-8 fibres. The concentration of peptide herein is 20 mg·mL$^{-1}$ (sample 3).

This effect is likely to be related to the much higher concentration of peptide used in the electrospinning solution (20 mg·mL$^{-1}$) compared to the average concentration previously analysed during CD spectroscopy of $P_{11}$-X peptides (0.15 mg·mL$^{-1}$) as well as the presence of PCL. FTIR analysis of the PCL/$P_{11}$-8 solution (Figure 1B) was in agreement with the CD data, whereby a broad peak was observed at around 1650 cm$^{-1}$, confirming a random coil configuration,[48] due to the HFIP-triggered breakdown of hydrogen bonds along the peptide molecules.[31, 47] The ATR-FTIR spectra for electrospun PCL/$P_{11}$-8 fibres are shown in Figure 1C. The presence of the peak at 1630 cm$^{-1}$ along with peaks at 1617 cm$^{-1}$ and 1683 cm$^{-1}$ gives supporting evidence of the predominant anti parallel β-sheet conformation of peptide in the fibres.[6, 48] At the same time, a very small peak at 1650 cm$^{-1}$ is observed in the FTIR spectrum of peptide-encapsulated electrospun fibres, suggesting the presence of a non-assembled peptide fraction in the fibres. Thus, the random coiled $P_{11}$-8 present in the PCL spinning solution is likely to self-assemble during electrospinning to yield predominant β-sheet conformation in the collected fibres. This is the result of the critical peptide concentration (c*) being reached during electrospinning due to solvent evaporation.

**Fibre morphology.** Electrospinning of the commixed PCL and $P_{11}$-8 peptide (Figure 2, C-H) induced detectable changes in fibre morphology as compared to spinning solutions based on PCL only (Figure 2, A-B). PCL fibres were homogenously distributed in terms of shape and diameter. In contrast, the webs containing both PCL and $P_{11}$-8 consisted of two superimposed fibre networks of submicron (100-700 nm) and nanofibres (10-100 nm). The fibre diameter distribution of PCL and PCL/$P_{11}$-8 (sample 1 and 3) is shown in Figure 3A and B. Note that the overall range of fibre diameters was greater for PCL/$P_{11}$-8 compared to the PCL control. In addition to the wide range of fibre diameters and the two distinct diameter distributions in fibres containing $P_{11}$-8, there is a decrease in the mean of submicron fibre region (Mean= 296.4 nm; Standard Error: 6.2 nm), compared to the PCL control (Mean = 386.9 nm; Standard Error: 5.6 nm). This observed shift can be attributed to an increase in conductivity of the polymer solution due to the peptide content ($P_{11}$-8 with +2 net charge), and therefore a likely increase in surface charge density of the jet during spinning, which causes a decrease in fibre diameter.[52, 53]

By increasing the concentration of $P_{11}$-8 in PCL solution to 10, 20 and 40 mg·mL$^{-1}$, the extent of the nanofibrous architecture present in the webs progressively increased (Figure 2), providing a convenient way of customising the nonwoven structure at the nanoscale. The nanofibres with diameter of 10-100 nm connected to the surrounding submicron parent fibres (Figure 2) were comparable in dimensions to $P_{11}$-8 fibrils reported in self-assembled gels.[2, 6] Moreover, some regions (e.g. in Figure 2C and E) revealed rope-like fibres that are consistent with aggregation of single β-sheet-ordered peptide tapes[48] or double twisted fibres at the submicron scale resembling the molecular orientation of single β-sheet tapes stacking face to face to form twisted fibrils.[3, 6]

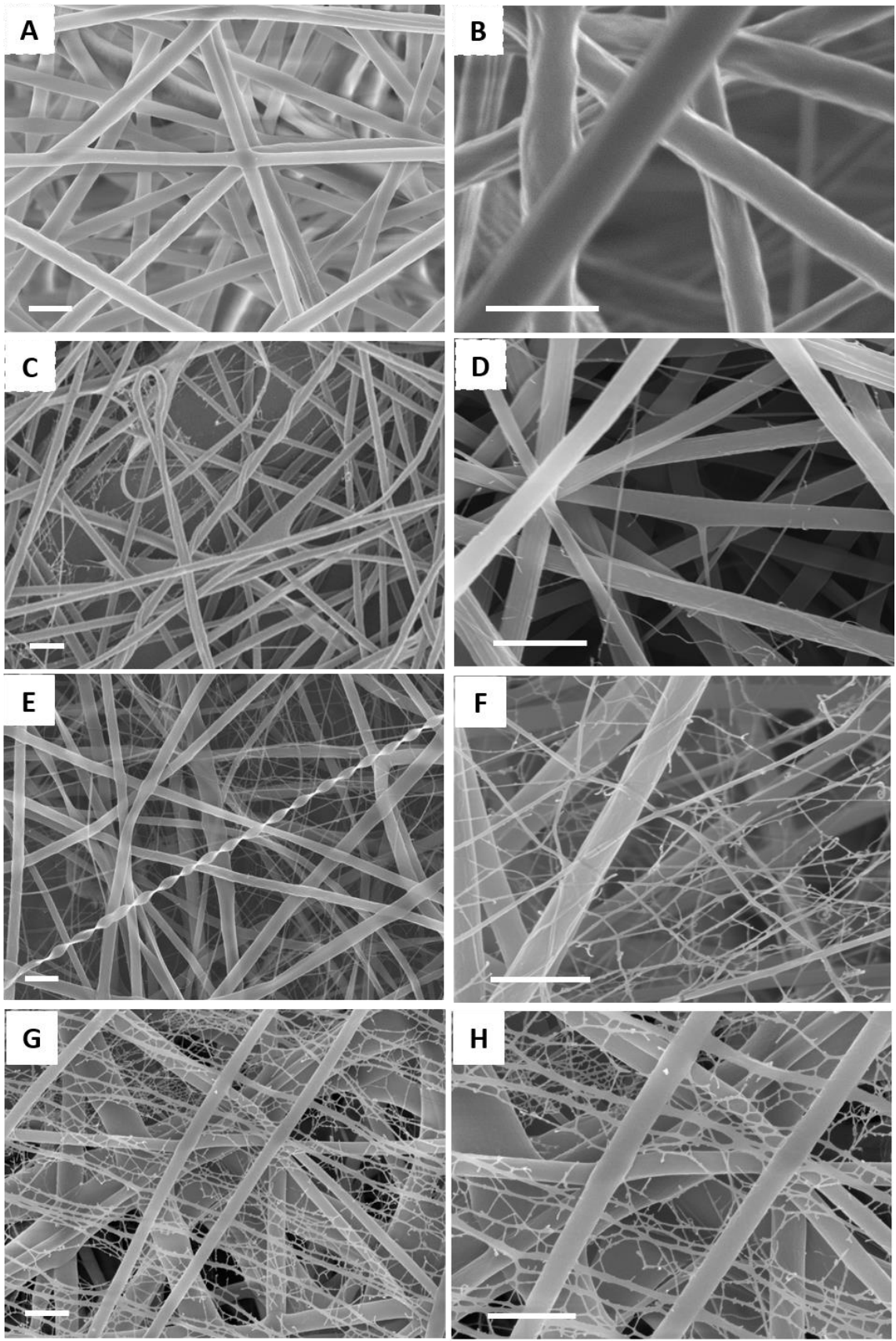

**Figure 2.** Scanning Electron Micrographs of samples 1 (A, B), 2 (C, D), 3 (E, F) and 4 (G, H). The scale bar in all images is 1 micron. By increasing the $P_{11}$-8 concentration in the electrospinning solution the degree of secondary superimposed nanofibre network (10-100 nm) formation increases amongst the submicron fibres (100-700 nm).

Based on these SEM observations and previous CD and FTIR spectra, it is likely that peptide assembly (from an unordered conformation to β-sheet) occurs during electrospinning of the solution by reaching the critical peptide concentration (c*) as a result of solvent evaporation. Ultimately further increase in peptide concentration triggers peptide monomers to self-assemble into structures of higher hierarchical assembly such as fibrils and bundled fibrils (fibres), following complete solvent removal in collected fibres. Such "spider-net" nanofibre architecture among submicron fibres has previously been reported by Pant et al.[54] for nylon-6 nanofibres as a result of hydrogen bond formation during fibre production. During electrospinning of peptides in the present study, renaturation of hydrogen bonds associated with peptide molecules, previously broken down by the presence of HFIP in solution,[55] could potentially promote the formation of a fibrillar network. Moreover the formation of electrospun spider-net like nanofibres has been related to an increase in conductivity of the spinning solution elsewhere caused by, for example, the addition of salts or secondary interactions between neighbouring molecules.[54, 56-62] In this study, $P_{11}$-8 peptide includes arginine residue at position 3, ornithine ($-(CH_2)_3NH_2$) residue at positions 5 and 7 and a glutamic acid unit at position 9 (Figure 3C). The side chains of arginine and ornithine are positively charged, whilst glutamic acid is negatively charged in solution resulting in an overall +2 net electrostatic charge.[6, 8] It can theoretically be assumed that by addition of $P_{11}$-8 into pure PCL solution, the electrical conductivity of the solution can increase and this has been shown in previous studies [45]. Therefore the progressive increase in the extent of nanofibre architecture (10-100 nm) with increasing peptide concentration is likely to be attributed to secondary repulsion effects between peptide monomers in neighbouring charged groups.

**Peptide distribution within the fibrous structure.** To verify the presence of peptide within the fibres, Energy Dispersive X-Ray Spectroscopy (EDX) was applied in conjunction with SEM and TEM. An example of the SEM-EDX spectra of the PCL fibres (sample 1) and PCL/$P_{11}$-8 at medium peptide concentration (sample 3) are shown in Figure 4A and B. The nitrogen peak, which can be attributed solely to the peptide, is observed in the PCL/P$_{11}$-8 spectrum with weight percentage of 8 ± 3%, compared to 0% nitrogen in the PCL control fibres. The EDX was also performed on sample 2, which was prepared with the lowest concentration of peptide (peptide concentration: 10 mg.mL$^{-1}$). The results did not indicate any detectable nitrogen content. This is in line with the fact that EDX cannot detect the presence of light elements (such as nitrogen) at concentration lower than 8 wt%.

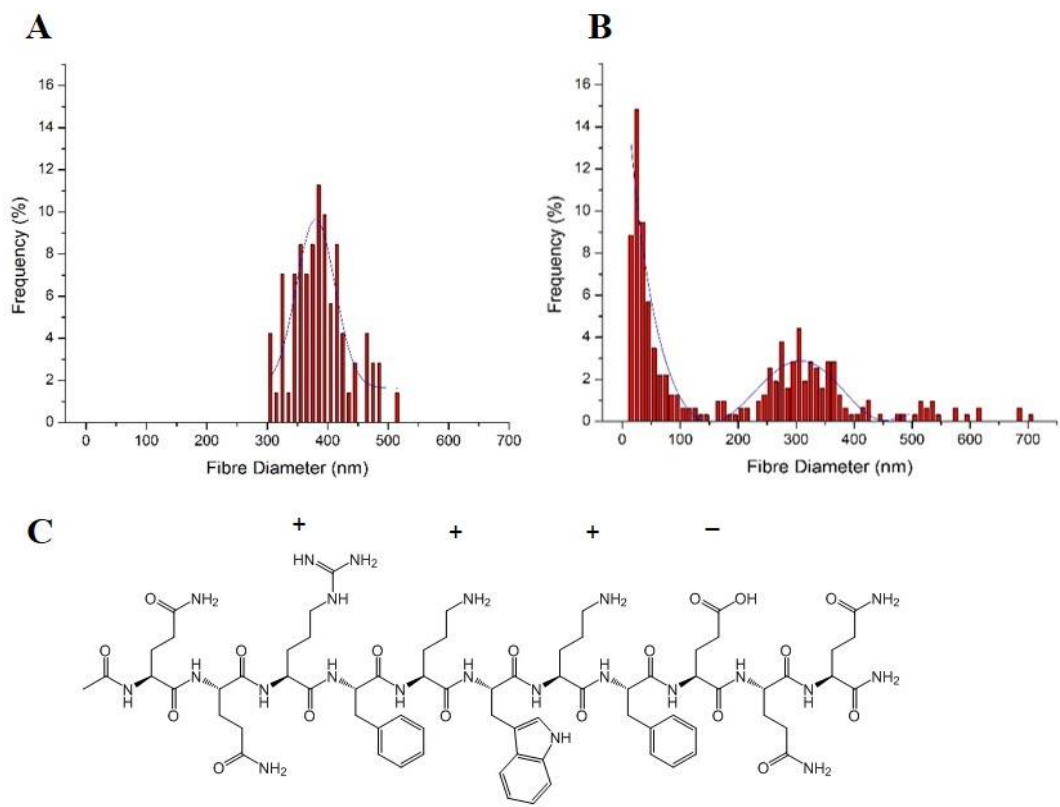

**Figure 3.** (A-B): Fibre diameter distributions of (A) samples 1 and (B) 3. (C) Peptide primary structure of P$_{11}$-8 carrying a +2 net electrostatic charge.

In addition to SEM-EDX, high resolution TEM images of the nano- and micro-scale fibres in the webs were obtained. Figure 4C shows a typical submicron fibre (diameter around 300 nm) and a single connected nanofibre (diameter around 50 nm) in sample 3 and EDX spectroscopy was applied to each of these fibres and confirmed nitrogen contents of 6 ± 1% and 4 ± 1%, respectively. The statistically insignificant variation of peptide-related nitrogen content on both fibrous networks only verifies that there is peptide present in both thick and thin fibres. The fact that TEM-EDX was carried out in 100% vacuum mode excludes the possibility of detecting environmental nitrogen residue. These observations are indicative of the incorporation of peptide within both the larger submicron (100-700 nm) and nanofibre architecture (10-100 nm), providing evidence of the absence of phase separation between PCL and the peptide during electrospinning.

CLSM micrographs of electrospun sample containing fluorescent peptides (sample 5) are shown in Figure 4D. No fluorescence background was detected in the PCL control in contrast to the PCL/$P_{11}$-8 sample where individual fibres were clearly visible confirming incorporation of $P_{11}$-8 throughout the entire fibrous structure, in agreement with the SEM and TEM elemental analysis.

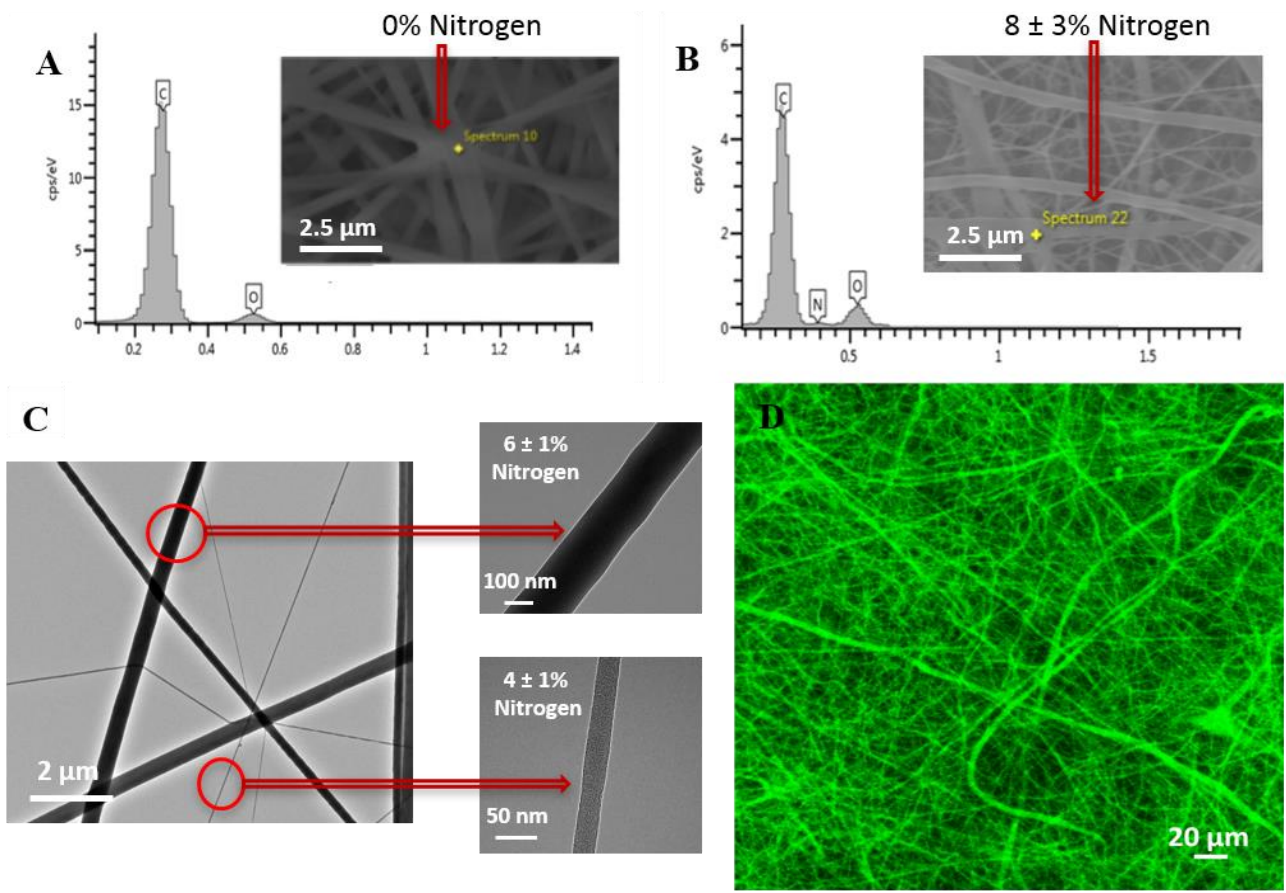

**Figure 4.** SEM/EDX analysis of electrospun fibres (A) sample 1 showing no nitrogen content and (B) sample 3 showing 8 ± 3% nitrogen content. (C) TEM/EDX analysis of sample 3; both submicron fibres and nanofibres were observed, whereby nitrogen content could be quantified. (D) Confocal laser scanning micrograph of fluorescently labelled electrospun fibres (sample 5) confirming the incorporation of peptide throughout the fibres.

**Surface wettability.** Given that water represents most of the weight fraction of the extracellular matrix of biological tissues, the contact angles of electrospun webs made of PCL and PCL/$P_{11}$-8 were measured via application of either deionised water or DMEM. Water and DMEM droplets on the surface of the webs (sample 1, 2 and 3) are shown in Figure 5 (A-F) at time = 0 s.

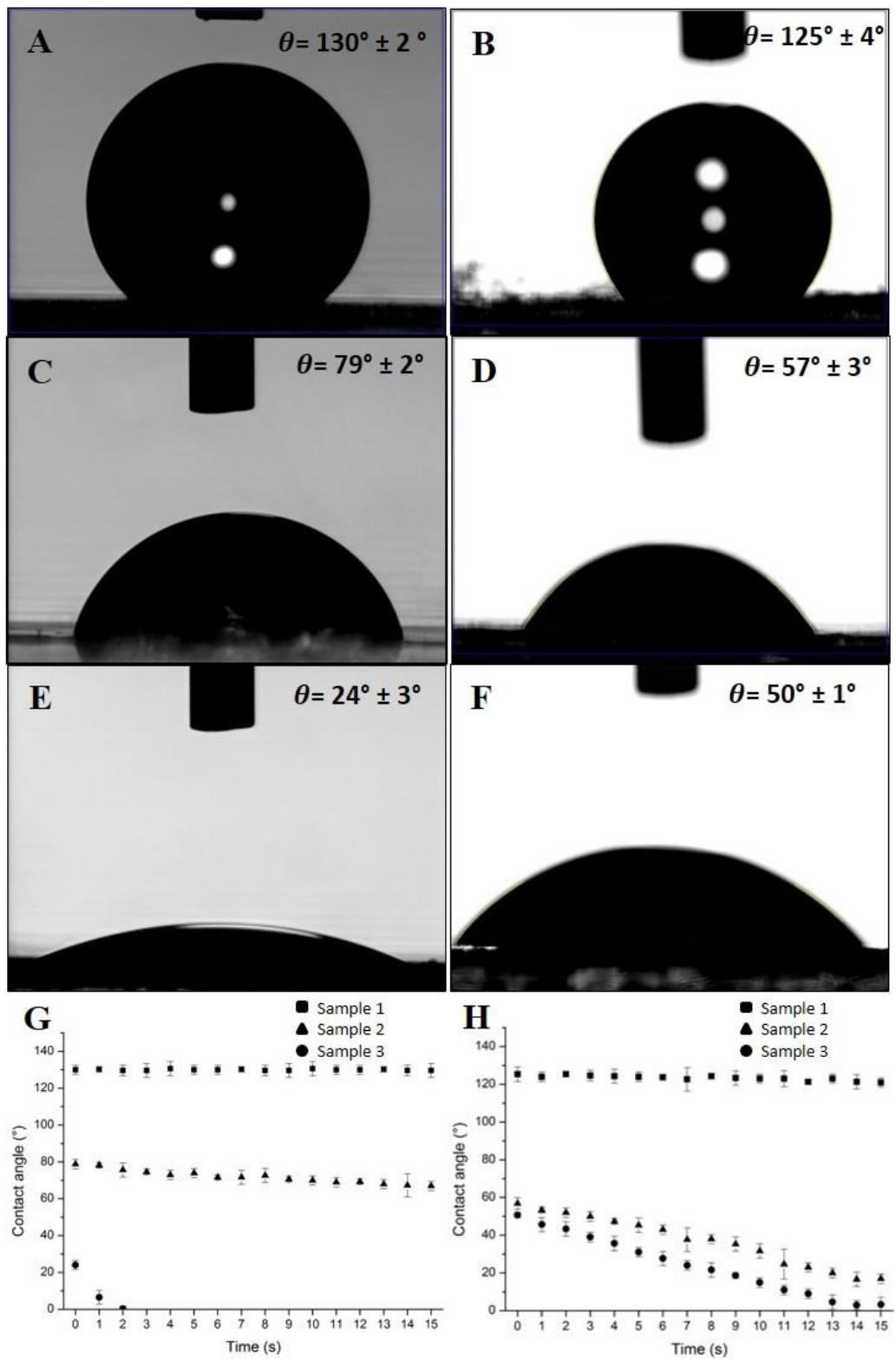

**Figure 5.** Initial contact angle of electrospun fibre webs following application of a droplet of either deionised water (A, C, E) or DMEM (B, D, F) on sample 1 (A, B), 2 (C, D) and 3 (E, F). Dynamic contact angle measurements on electrospun fibre webs for 15 s, droplet of both deionised water (G) and DMEM (H).

The contact angles on sample 4 could not be imaged and measured because of the rapid penetration of the droplet into the structure due to its hydrophilicity. Contact angles greater than 90° were observed on electrospun PCL, whilst decreased contact angle values were measured as $P_{11}$-8 concentration (0-20 mg·mL$^{-1}$) increased. This effect may be beneficial in order to promote cell adhesion and proliferation in PCL-based materials.[63] To improve the characterisation of wettability analysis in such porous samples, the rate of wetting was determined using goniometry, wherein several contact angles are measured as a function of time (duration of 15s).[64] Figure 5G and H show the dynamic contact angle for water and DMEM on the surfaces of sample 1, 2 and 3. Interestingly, the rate of wettability was found to progressively increase as the concentration of $P_{11}$-8 increased from 0 to 20 mg·mL$^{-1}$. The surface of electrospun webs comprises solid fibres and pores such that the resulting discontinuities can also potentially influence wetting behaviour. However, when films were made from the same electrospinning solutions and dried for 48 h at room temperature under vacuum, the observed contact angles were found to follow the same trend as the electrospun samples. As with the electrospun samples, the hydrophilicity improved with increasing peptide concentration from 0 up to 20 mg. mL$^{-1}$. This corresponded with contact angles ranging from $\theta = 79°$ to $\theta = 18°$. The data are shown in the supporting information, Figure S1.

**Direct cytotoxicity assay.** The cell response of L929 mouse fibroblasts when cultured in contact with sample 3 and sample 1 (negative control) fibre webs was assessed. Two typical images showing L929 cells in contact with both samples are shown in Figure 6A and B. A spread-like morphology was observed in cells cultured in direct contact with samples with and without $P_{11}$-8, such that fibroblasts proliferated up to and in contact with both the negative control and $P_{11}$-8-containing samples (dark areas are the edges of the webs). These morphologies can be verified by comparing with the morphology of cells that have been cultured in DMEM only as a control (data are provided in supplementary information Figure S2). These microscopic observations were supplemented by a MTT metabolic assay in order to quantify cell viability. MTT is metabolically reduced in viable cells to a blue-violet insoluble formazan. This assay (Figure 6C) revealed a comparable number of viable cells in contact with PCL/$P_{11}$-8 sample and the PCL and DMEM controls, whilst a significantly lower number of viable cells is detected in DMSO (positive control). Therefore, no evidence of a cytotoxic response could be discerned in the PCL/$P_{11}$-8 fibre web samples.

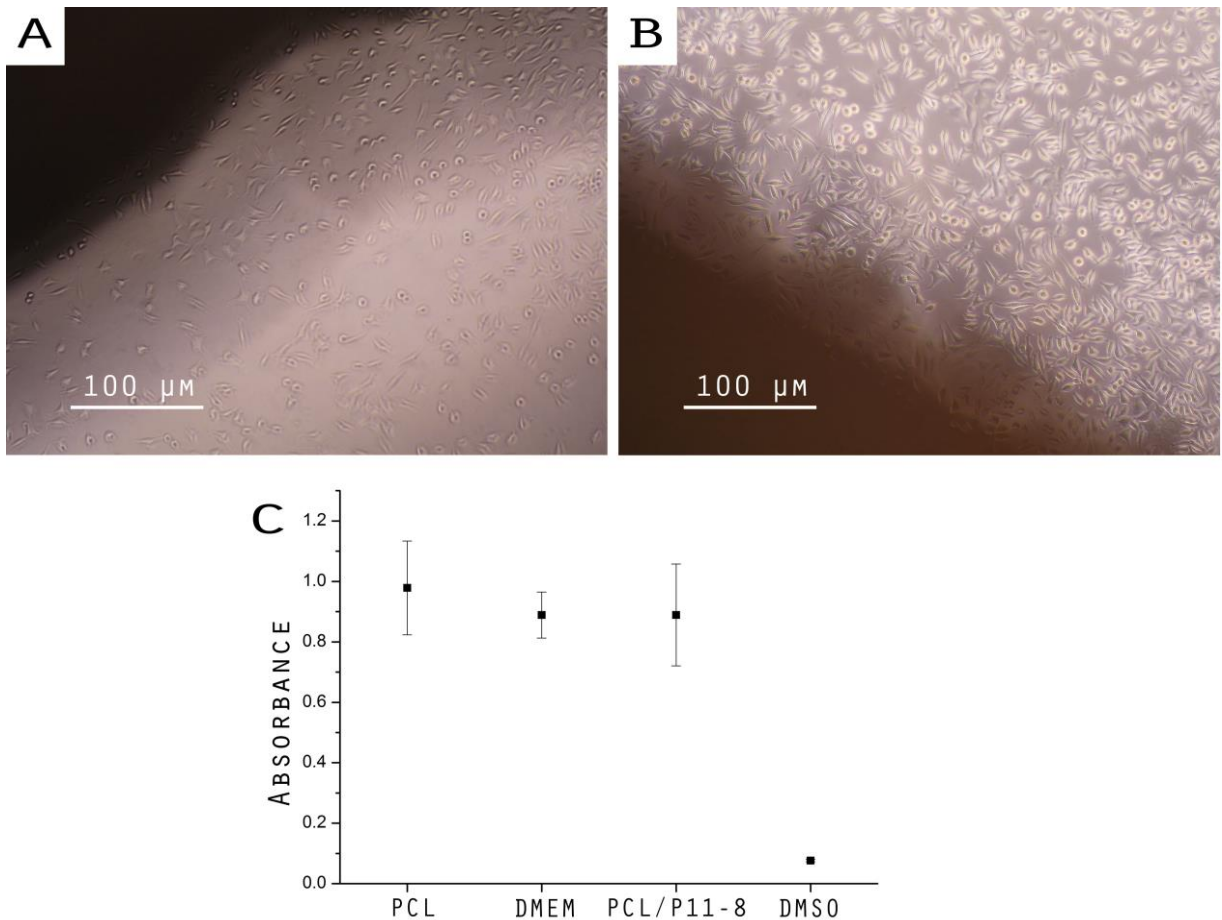

**Figure 6.** Light microscopy images of L929 cells in contact with sample 1 (A) and 3 (B). (C) OD absorption of samples at 570-650 nm correlated to the number of viable cells.

**Conclusions**

One-step assembly of a multiscale fibrous network containing micro-, nano- and *in-situ* molecular self-assembled $P_{11}$-8 peptide fibres was successfully demonstrated by electrospinning of a PCL-monomeric $P_{11}$-8 commixture. SEM, TEM and CLSM revealed a bimodal fibre diameter distribution of superimposed nano- and microscale fibre networks, both of which contain $P_{11}$-8 peptide. Analysis of the spinning solution and as-spun fibres by CD and FTIR revealed a switch from monomeric to predominant β-sheet peptide conformation, confirming that electrospinning process was able to trigger the molecular self-assembly mechanism and induce nanofibre formation. The hydrogen bonding-mediated self-assembly of $P_{11}$-8 during electrospinning, and the transition from a solution to fibre state, is most likely related to the rapid solvent evaporation and the resultant marked increase in the peptide concentration within the fibres (above c*). Although it is reasonable to assume that most of the solvent evaporation-driven self-assembly occurs via the solvent jet from the nozzle tip to the collector, it is possible that changes in peptide conformation continue if residual solvent is present in the deposited fibres. Adjusting the peptide concentration in the electrospinning solution was found to be a useful means of systematically customising the internal nanoscale structure of the webs, providing a convenient way of controlling nonwoven architecture at the nanoscale. The wide range of fibre diameters in samples containing $P_{11}$-8 may be beneficial in applications such as drug delivery where staggered release of multiple drugs is desirable. The combined micro- and nanofibrous architecture may also facilitate a controlled degradation profile in bicomponent scaffolds, since the nanofibres can be expected to degrade more quickly than the larger diameter fibres and therefore gradual degradation of the peptide component incorporated into submicron structures. Moreover, the presence of the biofunctional peptide among the submicron fibre pores could promote endogenous cell homing within the core of the material, which is crucial for functional tissue regeneration. Incorporation of $P_{11}$-8 in PCL enhanced hydrophilicity and proved to be well tolerated by mouse fibroblast cells. In light of these results, ongoing investigations are also focusing on the degradation kinetics and release profile of the peptide from the electrospun webs as well as the bioactivity of resulting materials for potential applications in bone/dental tissue repair.

**Acknowledgments**

The authors gratefully acknowledge financial support from the Alumni of the University of Leeds for the research scholarship awarded to RG and the support of the Clothworkers' Centre for Textile Materials Innovation for Healthcare. We are grateful to and wish to thank Dr. A. Aggeli for helpful discussions during the initial stages of this work.

**Supporting Information**

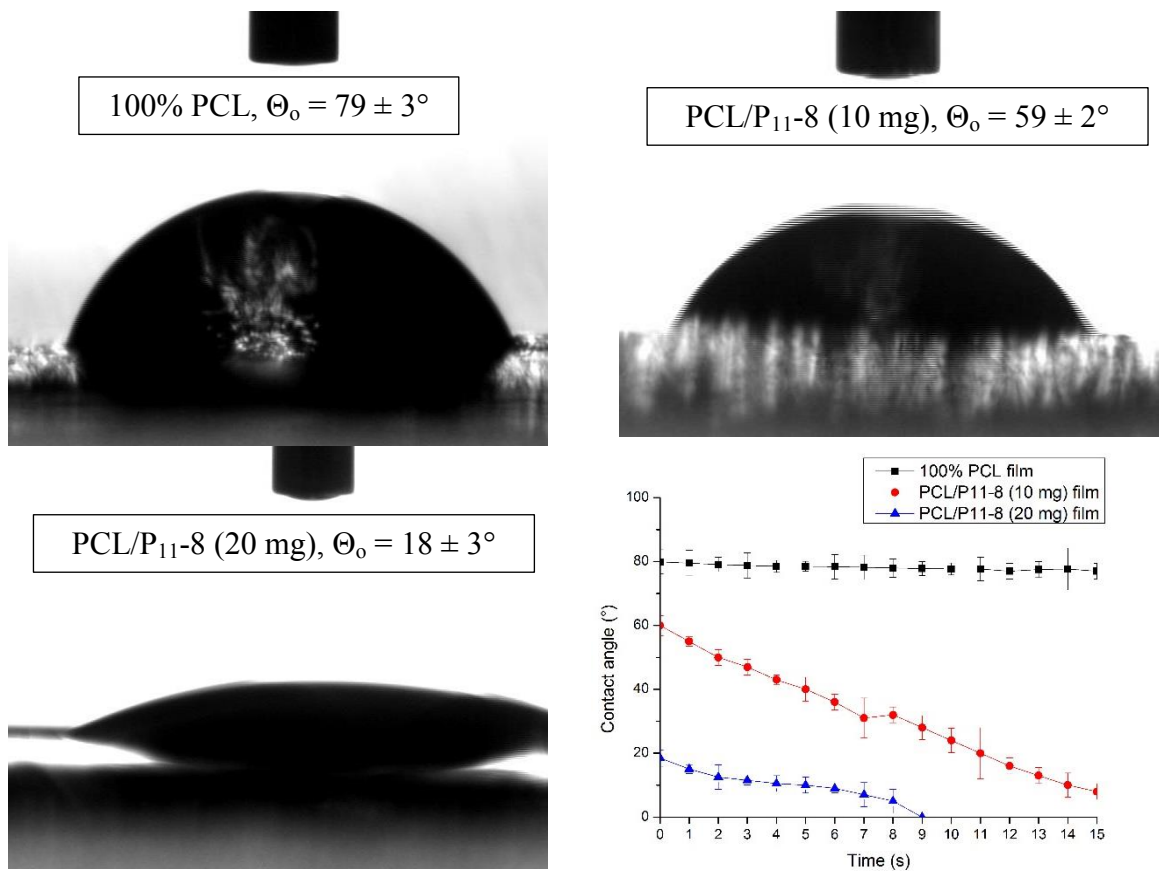

**Figure S1.** Contact angle of water on PCL and PCL/$P_{11}$-8 cast films showing enhanced hydrophilicity with increased peptide concentration.

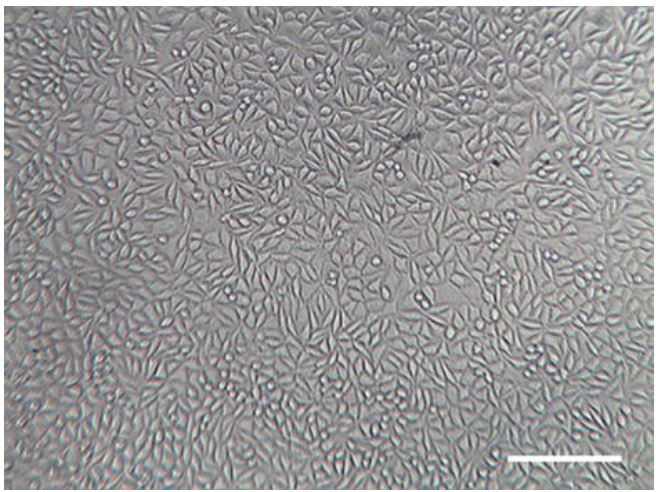

**Figure S2.** Light microscopy image of L929 cells cultured in DMEM as a control sample. Scale bar ~ 200 µm.